\documentclass[a4paper]{jpconf}
\usepackage{epsfig}
\usepackage{graphicx}
\begin{document}
\title{Implications of cosmic ray results for UHE neutrinos}

\author{Subir Sarkar}

\address{Department of Physics, University of Oxford, 1 Keble Road,
  Oxford OX1 3NP, UK}

\ead{s.sarkar@physics.ox.ac.uk}

\begin{abstract}
  Recent measurements of the spectrum and composition of ultrahigh
  energy cosmic rays suggest that their extragalactic sources may be
  accelerating heavy nuclei in addition to protons.  This can suppress
  the cosmogenic neutrino flux relative to the usual expectation for
  an all-proton composition. Cosmic neutrino detectors may therefore
  need to be even larger than currently planned but conversely they
  will also be able to provide valuable information concerning
  astrophysical accelerators. Moreover measurement of ultrahigh energy
  cosmic neutrino interactions can provide an unique probe of QCD
  dynamics at high parton density.
  \end{abstract}

\section{Introduction}

Cosmic rays have been detected with energies up to
$\sim3\times10^{20}$~eV. Ultra high energy neutrinos {\em must} also
be generated during their interactions with ambient matter and
radiation in the sources, and with intergalactic radiation backgrounds
during their propagation to Earth \cite{Halzen:2002pg}. The detection
of these neutrinos would enable unambiguous identification of the
sources, as well as probe new physics both in and beyond the Standard
Model \cite{Anchordoqui:2005is}.

In order to estimate the expected event rates in cosmic neutrino
detectors such as ANITA \cite{Barwick:2005hn}, HiRes
\cite{Bergman:2008at}, IceCube \cite{Klein:2008mw}, and the Pierre
Auger Observatory \cite{Roulet:2008jq}, it is thus essential to take
new data on ultra high energy cosmic rays (UHECRs) into account. There
has been significant recent progress in the field, in particular HiRes
\cite{Abbasi:2007sv} and Auger \cite{Abraham:2008ru} have established
that the energy spectrum is attenuated beyond $\sim 5 \times
10^{19}$~eV. This is indeed as expected if the primaries are protons
undergoing photopion interactions on the cosmic microwave background
(CMB) \cite{Greisen:1966jv,Zatsepin:1966jv}. Such interactions should
also give rise to a flux of high energy neutrinos
\cite{Berezinsky:1970xj} and this ``guaranteed cosmogenic flux'' is a
prime target for cosmic neutrino detectors. A somewhat higher (but
more model-dependent) flux of neutrinos should also be generated in
the sources of the cosmic rays through $pp$ and $p\gamma$ interactions
\cite{Waxman:1998yy}.

However it is not clear that the UHECRs are necessarily protons.
Astrophysical accelerators are expected to generate particles up to
a maximum energy which is proportional to their charge
\cite{Hillas:1985is} hence it would be less challenging for plausible
sources such as active galactic nuclei (AGN) to emit $\sim10^{20}$~eV
iron nuclei rather than protons. The correlation observed by Auger
between the arrival directions of UHECRs above $6 \times 10^{19}$~eV
and AGN within 75 Mpc \cite{Abraham:2007si} would seem to argue
against heavy nuclei as primaries since these ought to be
significantly deflected by intergalactic and galactic magnetic
fields. However UHECR nuclei will undergo photodisintegration on the
cosmic infrared background (CIB) with a energy loss length similar to
protons \cite{Stecker:1969fw} so the cosmic rays arriving at Earth
will be much lighter, thus reducing the impact of magnetic fields
especially in the Galaxy. Moreover intergalactic magnetic fields may
be weaker than is usually assumed --- observationally only upper
limits are known.

The chemical composition of UHECRs can in principle be inferred from
the development of the air showers they trigger on hitting the Earth's
atmosphere. For a given energy, the depth at which the shower reaches
maximum development, $X_{\rm max}$, is smaller for heavy nuclei than
for protons, and its average value increases logarithmically with
energy. There is however considerable scatter due to fluctuations
associated with the stochasticity of the first interaction and
moreover different (semi-empirical) simulation codes for air showers
make differing predictions for $X_{\rm max}$
\cite{Anchordoqui:2004xb}. Earlier data from HiRes had suggested that
the composition becomes light in the range $\sim
10^{18}-3\times10^{19}$~eV \cite{Abbasi:2004nz}. However recent
measurements by Auger \cite{Unger:2007mc} which reach somewhat higher
in energy indicate a gradual {\em decrease} in $X_{\rm max}$ above
$\sim 2 \times 10^{18}$~eV, implying increasing dominance by heavy
nuclei. This would argue against the interpretation of the `ankle' in
the energy spectrum at $\sim10^{19}$~eV as due to $e^+e^-$ energy
losses of extragalactic cosmic ray {\em protons} on the CMB
\cite{Berezinsky:2005cq}. The alternative explanation is that at this
energy the flatter spectrum of extragalactic cosmic rays dominates
over the falling galactic component, whereas in the former case the
transition must occur at a lower energy of $\sim 10^{18}$ eV (`second
ankle') and require fine-tuning between the two components to ensure a
smooth transition.

Hence it is necessary to determine the range of possible compositions
for the primary particles which is consistent with the energy spectrum
and $X_{\rm max}$ measured at Earth. To do this we must compute the
propagation of UHECR nuclei through the CIB to match our understanding
of the propagation of UHECR protons through the CMB and the generation
of the cosmogenic neutrino flux
\cite{Engel:2001hd,Fodor:2003ph}. Nuclei would undergo
photodisintegration (at a lower energy threshold than that for pion
production) and the secondary nucleons, if still sufficiently
energetic would then produce pions through the usual GZK process, also
the neutrons would undergo $\beta$-decay. Depending on the choice of
chemical composition and injected spectrum of the UHECRs, the
cosmogenic neutrino spectrum can in some cases be considerably
suppressed relative to that predicted for an all-proton composition.

\section{The cosmogenic neutrino flux}

The complex process of the photodisintegration of UHECR nuclei into
lighter nuclei and nucleons has been addressed using Monte Carlo
techniques by several authors
\cite{Yamamoto:2003tn,Hooper:2004jc,Ave:2004uj,Allard:2006mv,Hooper:2006tn,Anchordoqui:2007fi},
but it is useful to develop an {\em analytic} description of this
phenomenon \cite{Hooper:2008pm}.

\begin{figure}[!tbp]
\epsfig{file=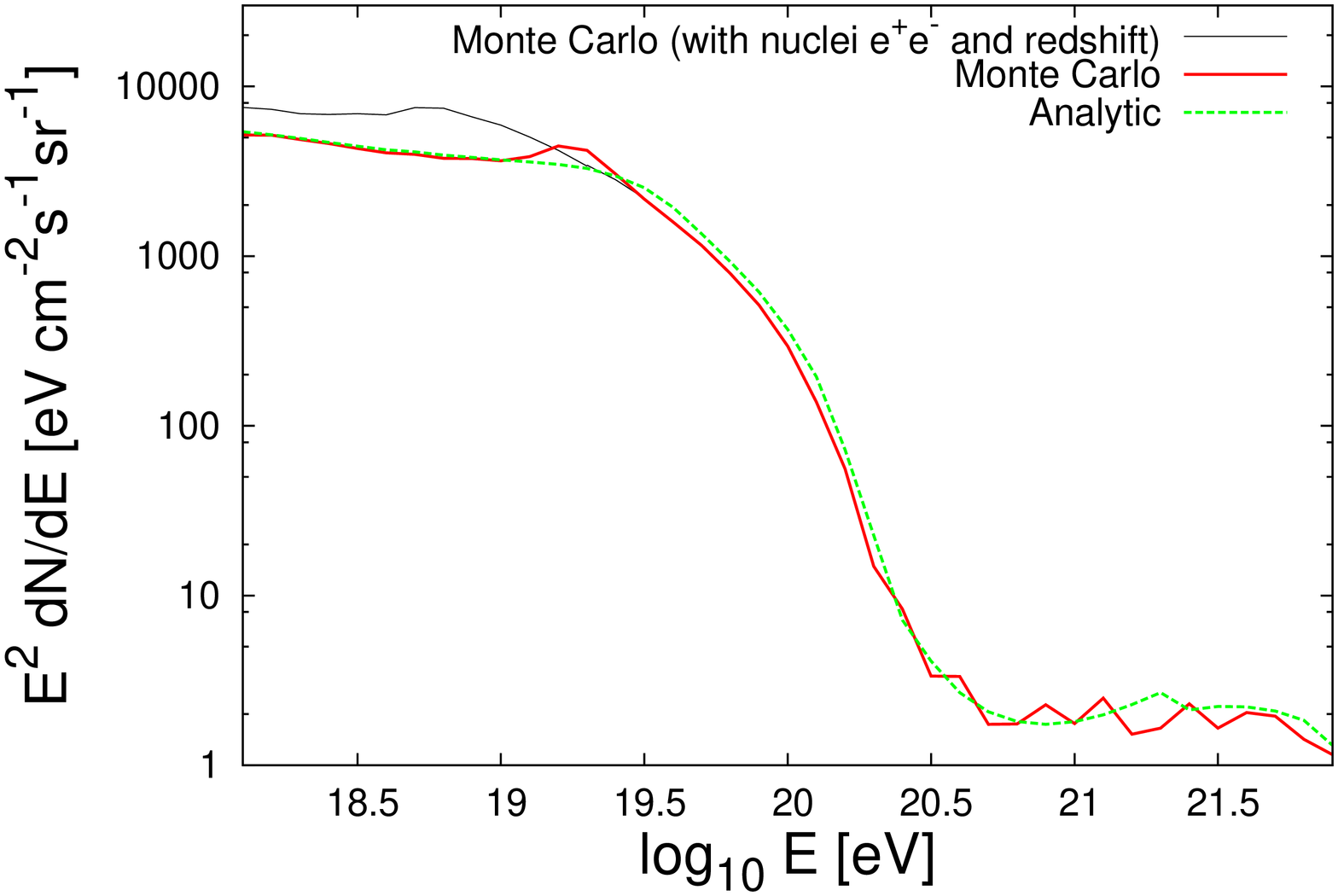,angle=-90,width=.5\textwidth}
\hfill
\epsfig{file=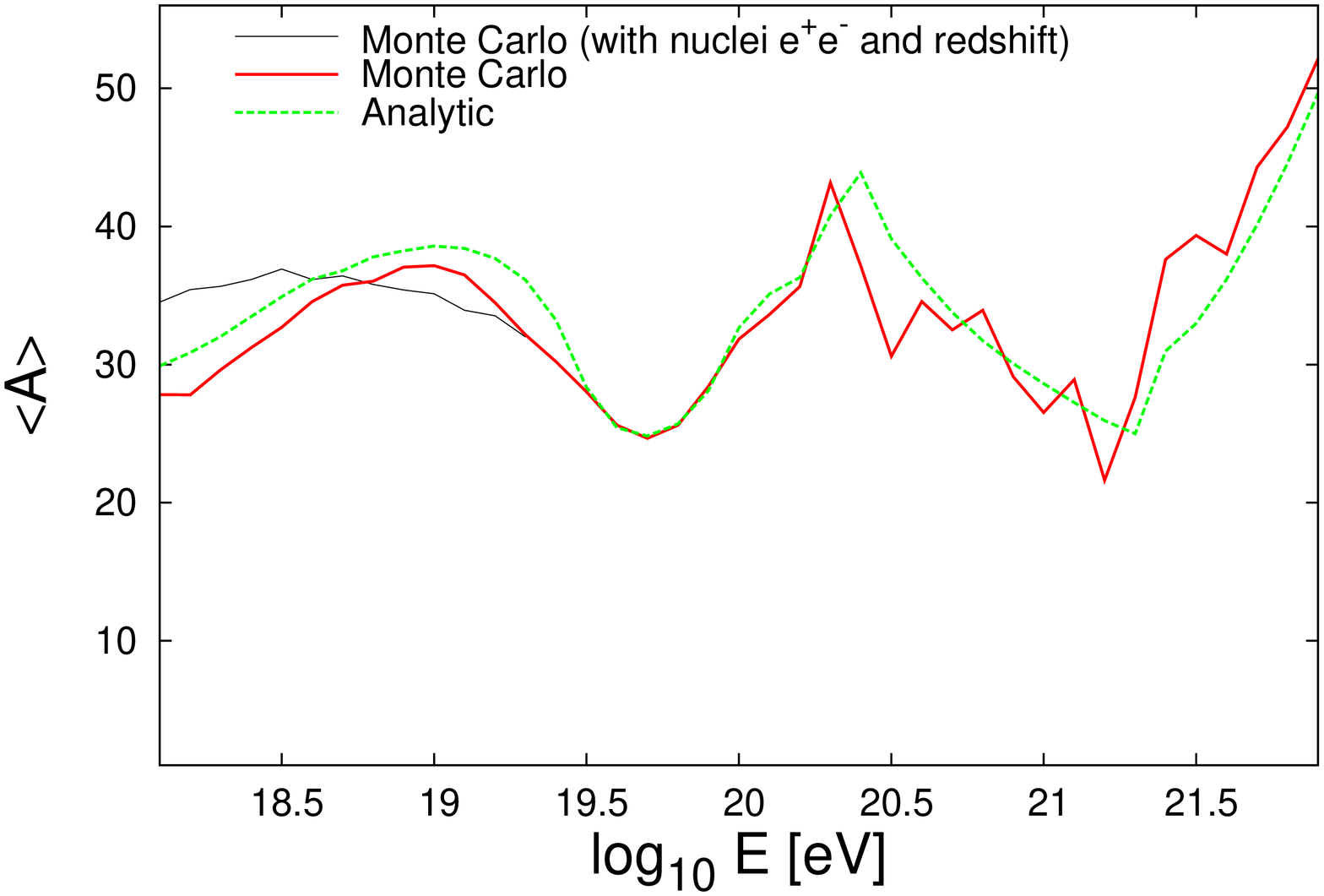,angle=-90,width=.5\textwidth} 
\caption{The energy spectrum (left) and average composition (right) at
  Earth calculated using both analytic and Monte Carlo techniques, for
  the case of iron nuclei injected by homogeneously distributed
  sources with $\mathrm{d}N/\mathrm{d}E \propto E^{-2}$ up to a
  maximum energy of $10^{22}$ eV \cite{Hooper:2008pm}.}
\label{Analytic_compare}
\end{figure}

\begin{figure}[!tbp]
\epsfig{file=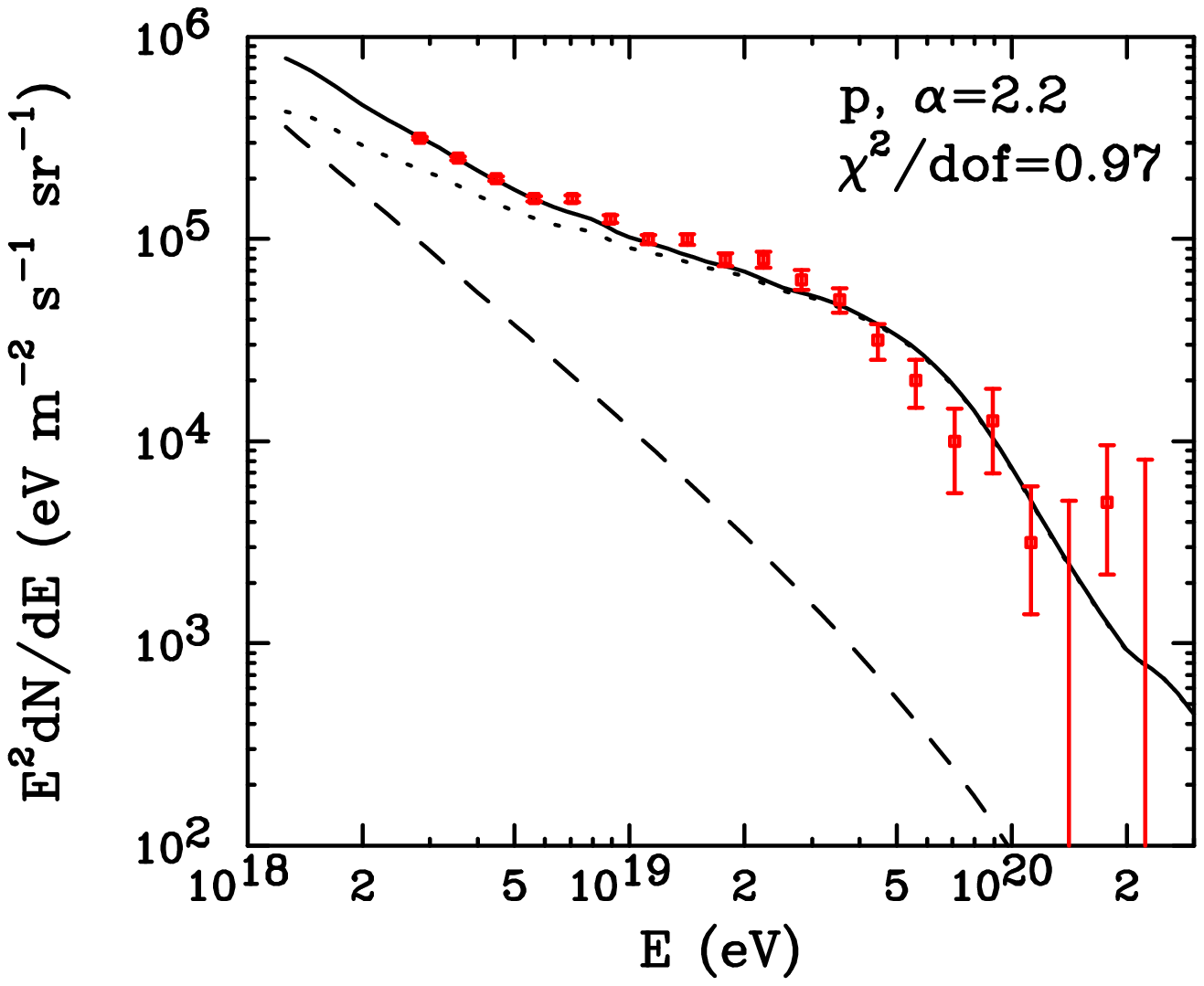,width=.45\columnwidth}
\hfill
\epsfig{file=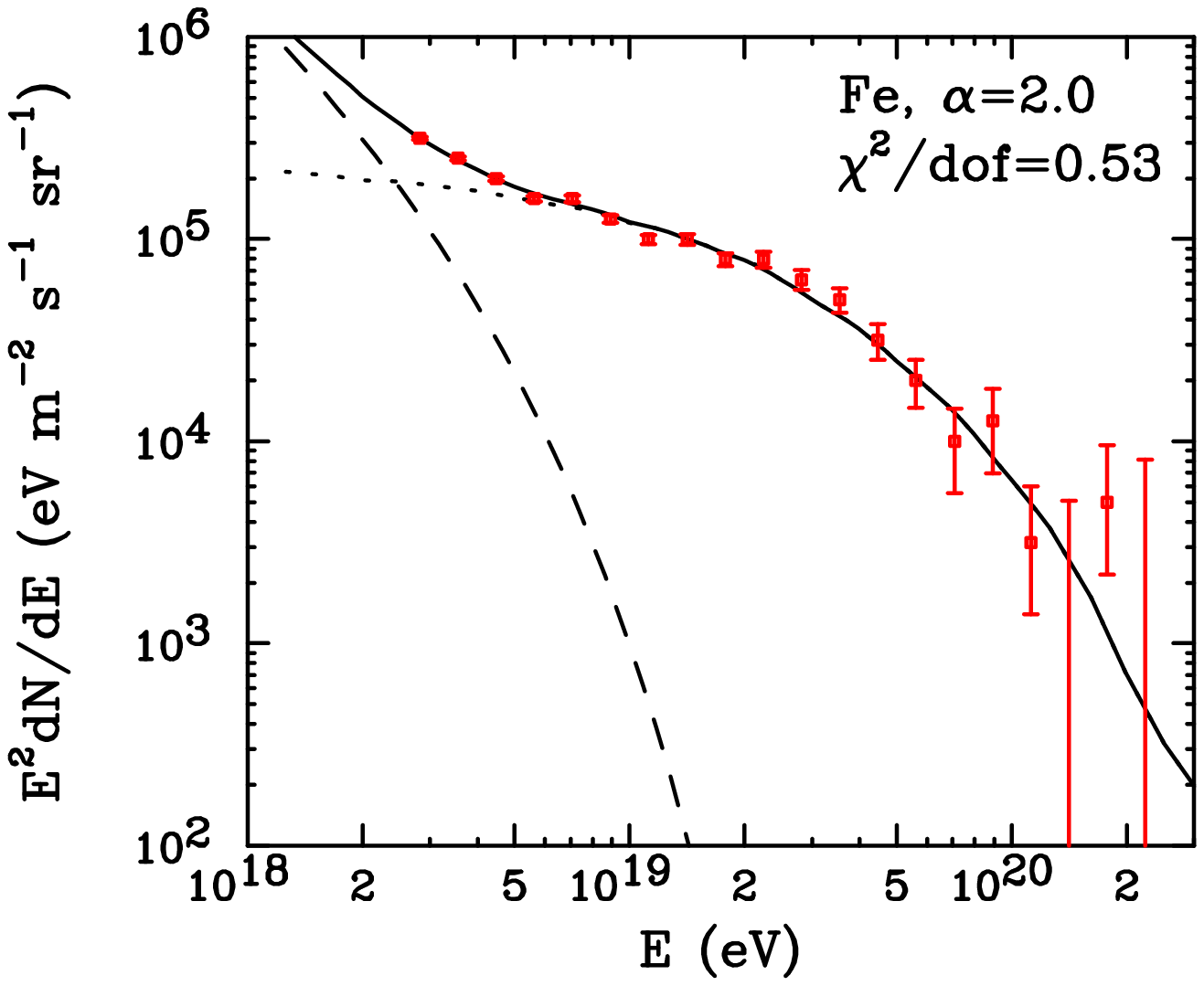,width=.45\columnwidth} 
\caption{The best fit spectrum for an all-proton and all-iron UHECR
  composition with injection power-law slopes of -2.2 and -2.0
  respectively --- the dotted, dashed and solid lines denote the
  assumed extragalactic, galactic and combined components
  \cite{Anchordoqui:2007fi}.}
\label{protonspec}
\end{figure}

\begin{figure}[!tbp]
\epsfig{file=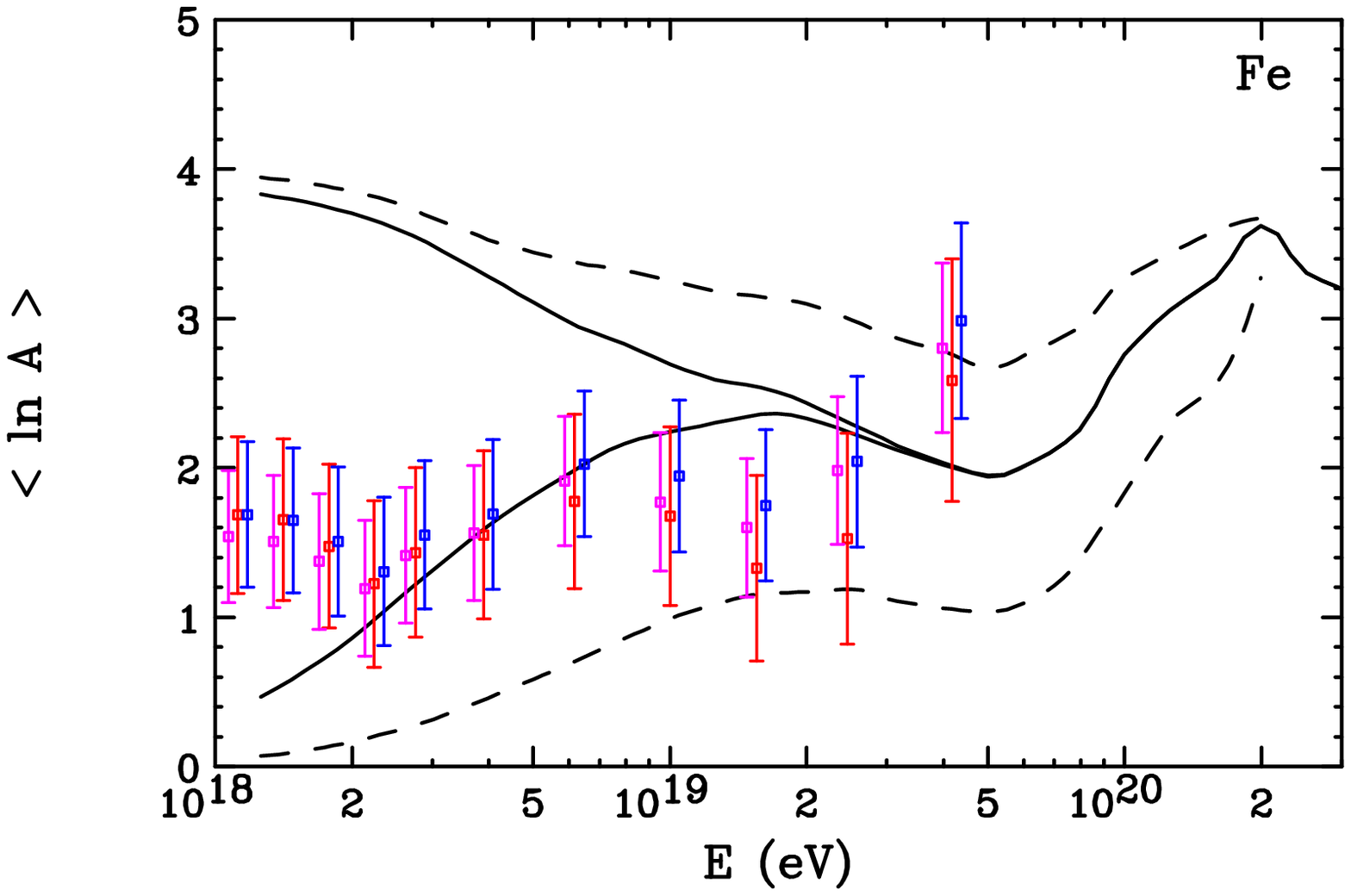,width=.45\columnwidth}
\hfill
\epsfig{file=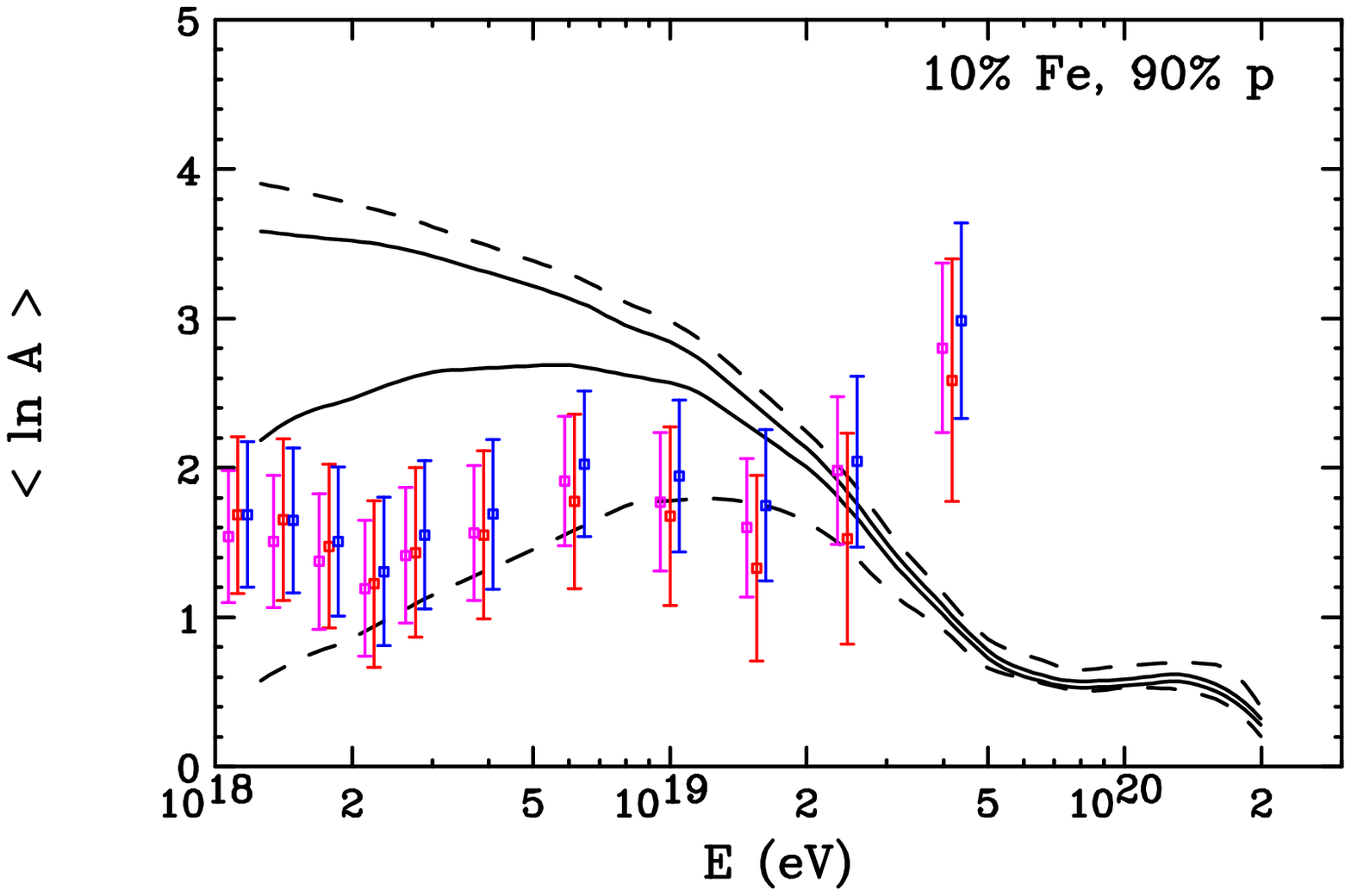,width=.45\columnwidth}
\caption{The composition at Earth when pure iron nuclei (left) or a
  mixture of protons and iron nuclei (right) are injected by the
  sources, if we require that after propagation the measured energy
  spectrum at Earth is consistent with the Auger data --- the
  broadening below $\sim 10^{19}$ eV results from possible variations
  in the composition of the galactic component while the dashed lines
  denote the 95\% c.l. range. The data points correspond to the
  $X_\mathrm{max}$ measurements by Auger interpreted according three
  different hadronic physics models: EPOS 1.6 (magenta), QGSJET-III
  (red) and SIBYLL 2.1 (blue), including both systematic and statistical
  errors \cite{Anchordoqui:2007fi}. }
\label{xmax}
\end{figure}

This has turned out to be very successful,
e.g. Figure~\ref{Analytic_compare} shows a comparison of our analytic
\cite{Hooper:2008pm} and Monte Carlo
\cite{Hooper:2004jc,Hooper:2006tn} results for the case when iron
nuclei are injected by the sources. This both validates and provides
valuable insights into the Monte Carlo results. Given the
uncertainties concerning the galactic-extragalactic transition (and
the composition of the galactic component), we find that the Auger
data concerning the energy spectrum \cite{Abraham:2008ru} as well as
the composition at Earth \cite{Unger:2007mc} can be satisfactorily
fitted with a wide range of nuclei being injected, as is illustrated
in Figures~\ref{protonspec} and \ref{xmax} \cite{Anchordoqui:2007fi} .

The corresponding cosmogenic neutrino flux is however very different
and can be suppressed significantly relative to the all-proton
case. However as shown in Figure~\ref{xmax}, the data is also
consistent with a proton-dominated spectrum with a small admixture of
heavy nuclei, in which case the cosmogenic flux will still yield of
${\cal O}(1)$ cosmogenic neutrino event per year in a kilometer-scale
neutrino telescope. With a bigger detector, it may even be possible to
constrain the composition at injection and the free parameters in the
calculation, e.g. the spectral slope and maximum energy to which
particles are accelerated as well as possible evolution of the number
density of sources with redshift which we have not considered
here. Note that while the observed UHECRs cannot come from very far
away because of the GZK energy losses, the universe remains
transparent even to such high energy neutrinos back to the
recombination epoch.

\begin{figure}[!tbp]
\epsfig{file=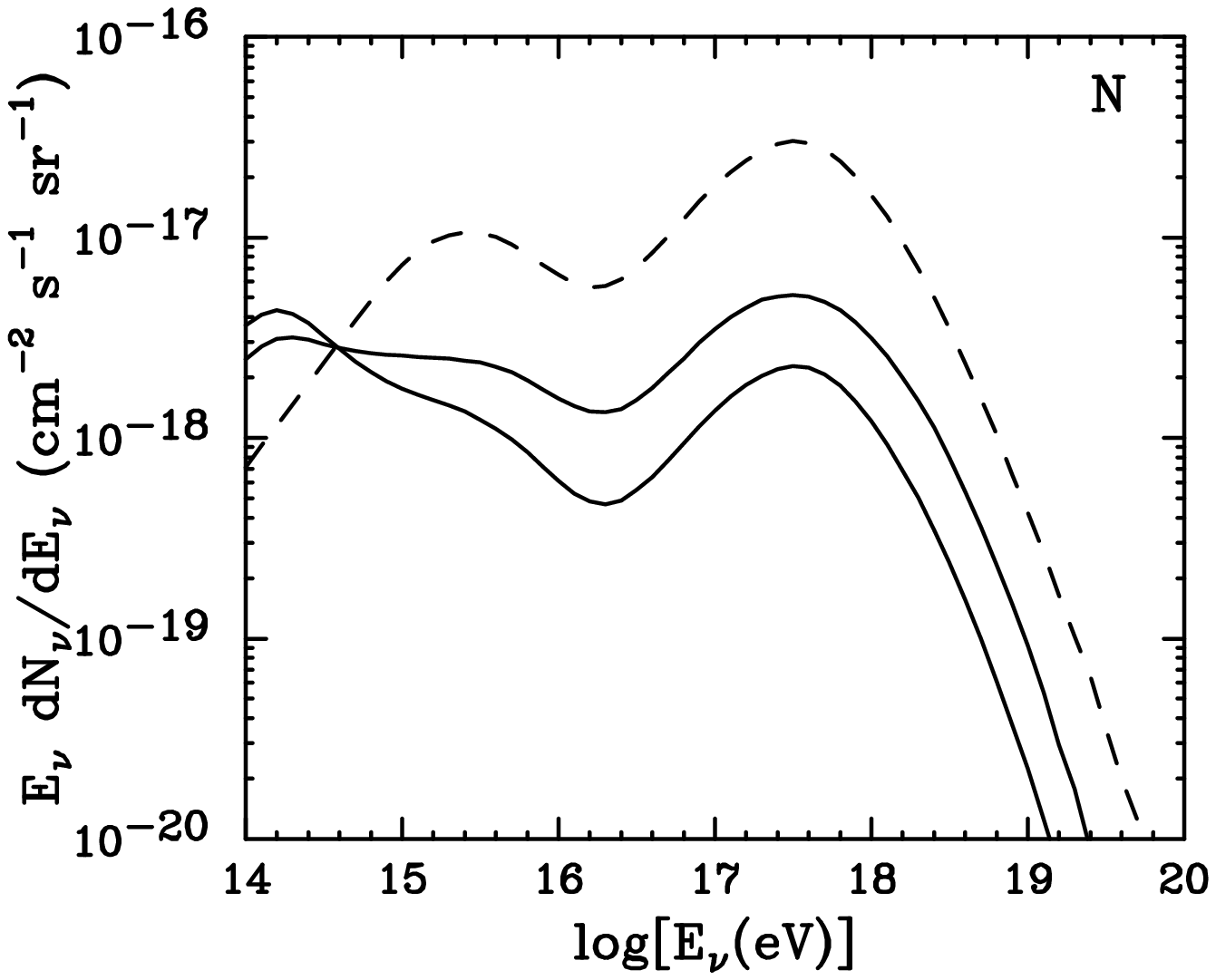,width=.45\textwidth}
\hfill
\epsfig{file=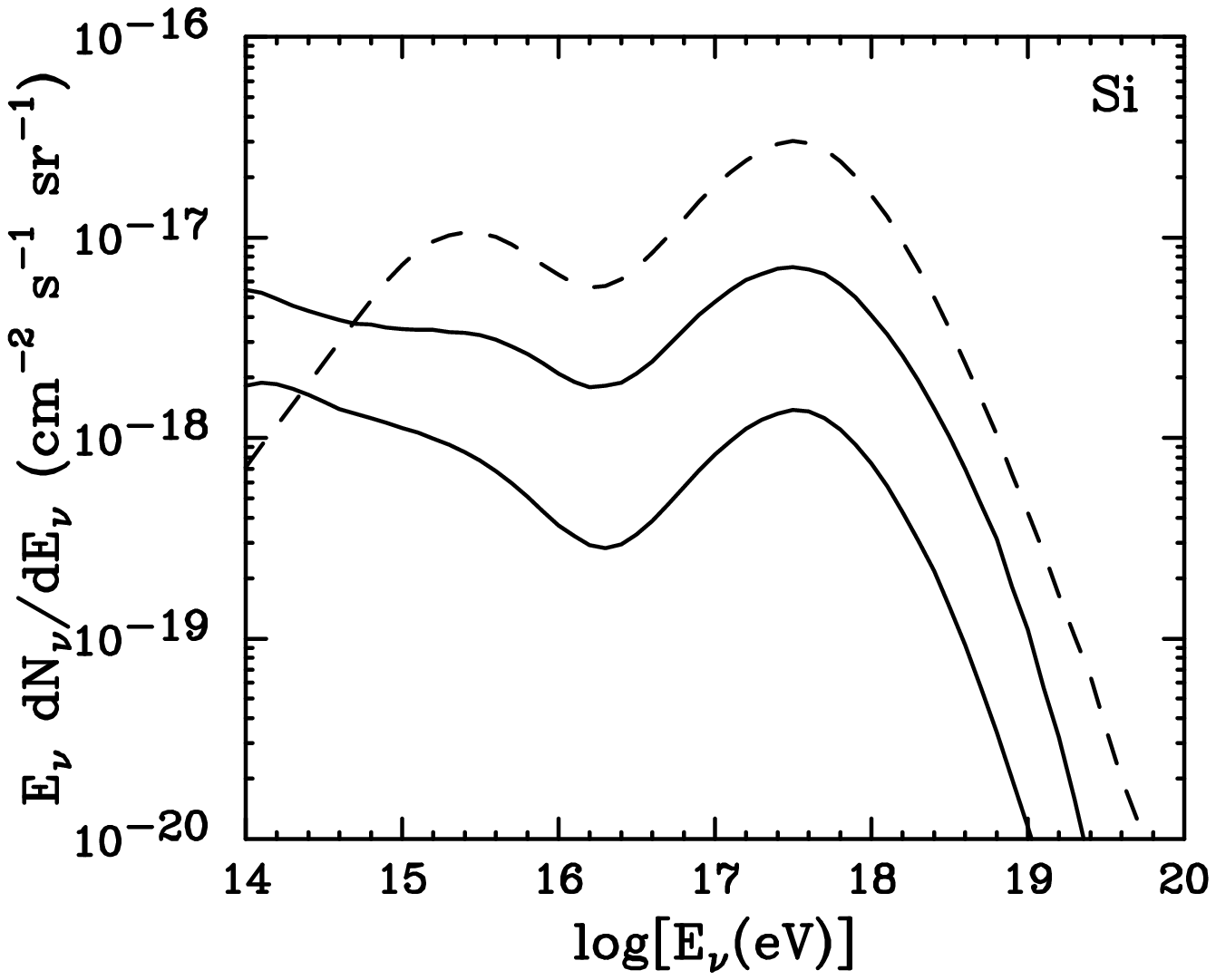,width=.45\textwidth}\\ 
\epsfig{file=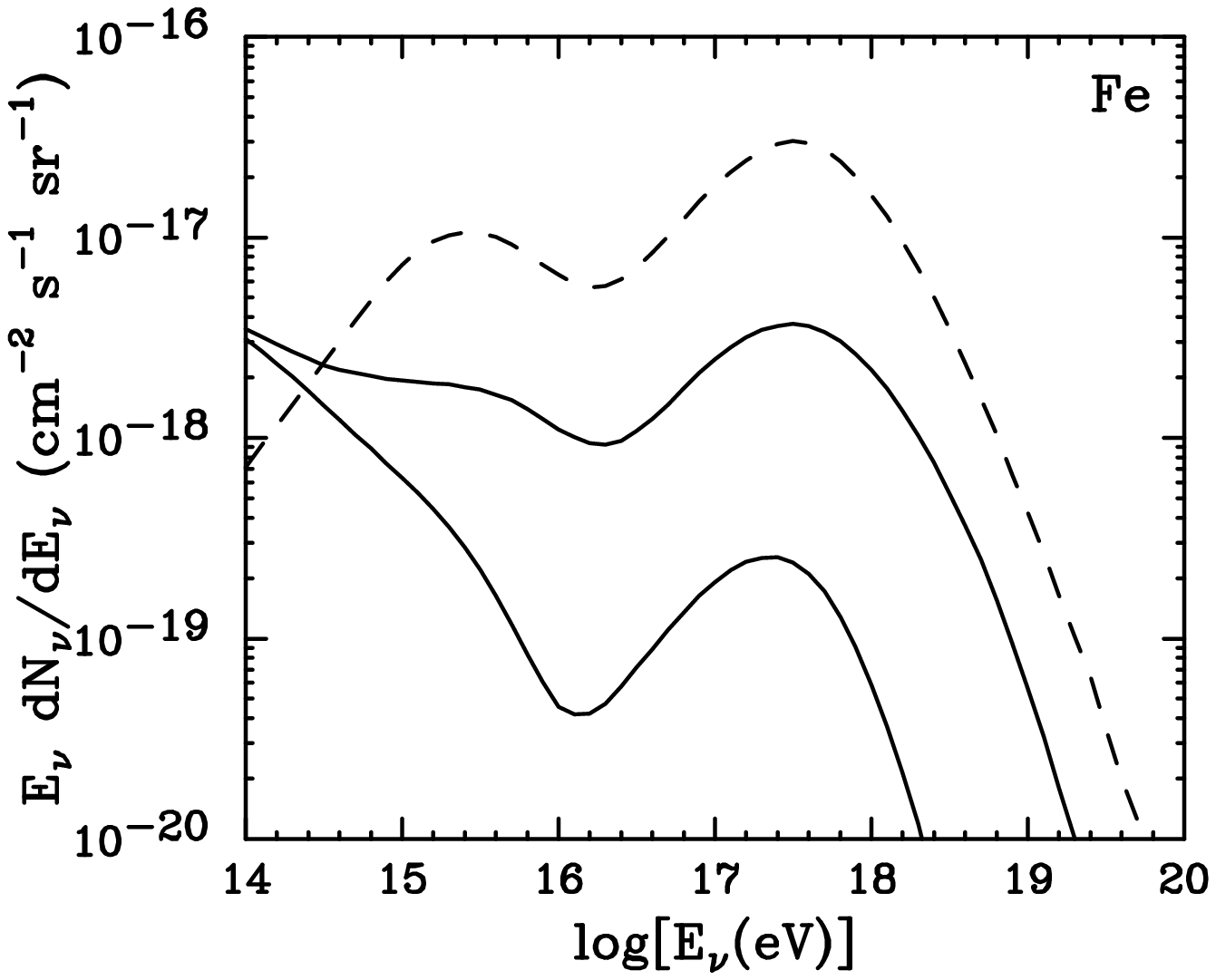,width=.45\textwidth}
\hfill
\epsfig{file=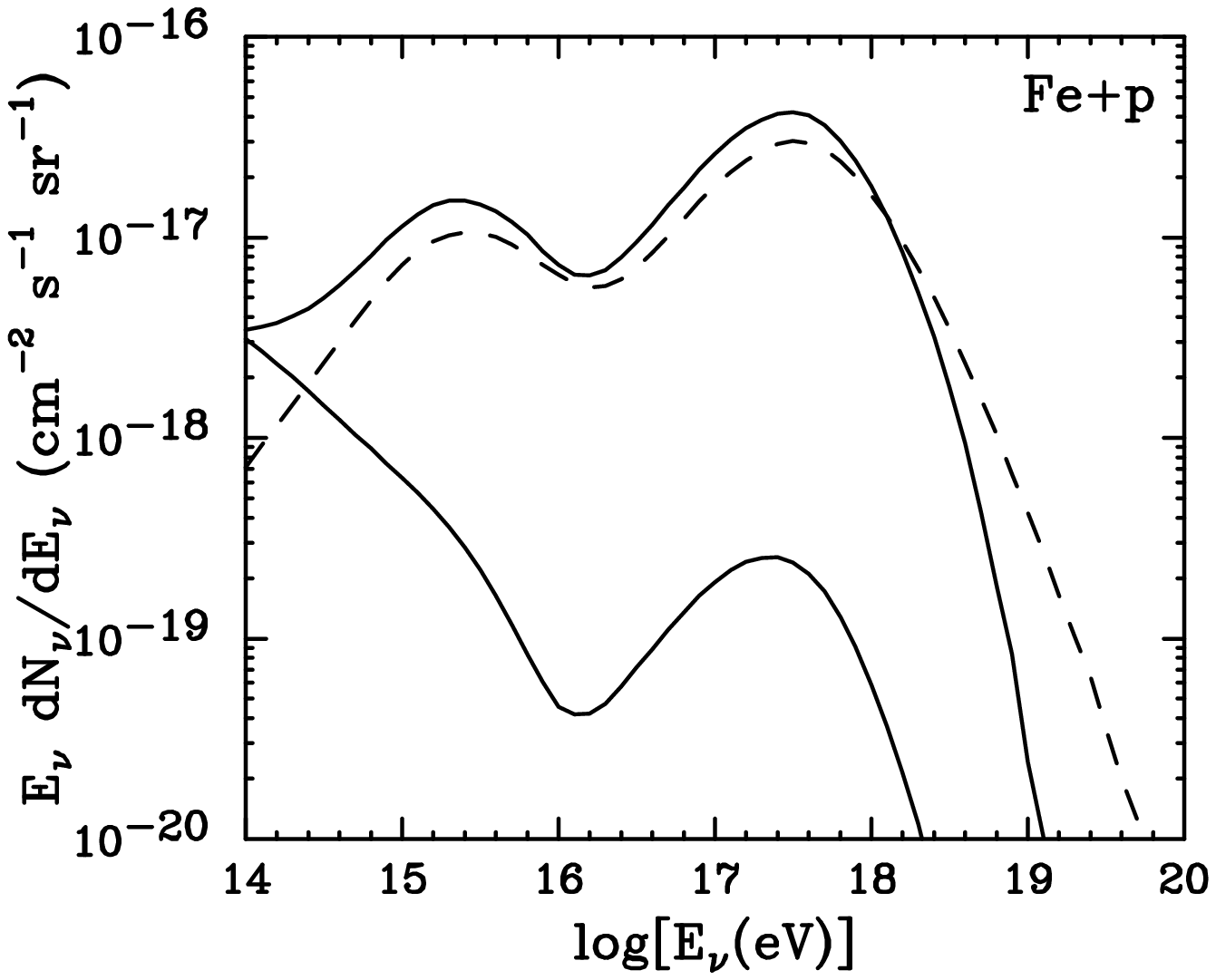,width=.45\textwidth}\\
\caption{The range of cosmogenic neutrino spectra for various injected
  chemical species which, after propagation, are consistent with the
  Auger spectrum and $X_{\rm{max}}$ measurements; the dashed curve is
  for an all-proton spectrum with power-law slope $\alpha=-2.2$ and
  $E_{\rm{max}} =10^{22}$ eV \cite{Anchordoqui:2007fi}.}
\label{neutrinospec}
\end{figure}

\section{Neutrinos from cosmic ray accelerators}

The neutrino flux expected from the extragalactic sources of cosmic
rays depends on how the particles are accelerated and the environment
in which this occurs. Assuming that the sources are `optically thin'
and normalising to the observed UHECR spectrum, an upper limit can be
placed on the diffuse flux \cite{Waxman:1998yy} which is only a little
higher than a plausible estimate for the actual flux based on the
known efficiency of $pp$ and $p\gamma$ interactions for producing
neutrinos. This estimate is of course significantly higher in the `low
cross-over' model for the galactic--extragalactic transition since the
sources must then put much more power into generating cosmic rays
\cite{Ahlers:2005sn}. Of course just like the cosmogenic flux, all
such estimates are sensitive to the assumed composition at
injection. We find that based on what is observationally known about
the environment in suggested sources, heavy nuclei are likely to be
completely photodissociated in $\gamma$-ray bursts but survive
unscathed in starburst galaxies, while the situation in AGN is
somewhere in-between \cite{Anchordoqui:2007tn}. With regard to
individual objects, the detection of correlations between UHECR
arrival directions and nearby AGN has inspired many estimates of the
expected neutrino flux, e.g. {\sl Centaurus A} may yield $0.4-0.6$
events/yr in IceCube \cite{Cuoco:2007qd}

\section{Neutrino interaction cross-sections}

\begin{figure}[t]
\centerline{
\epsfig{figure=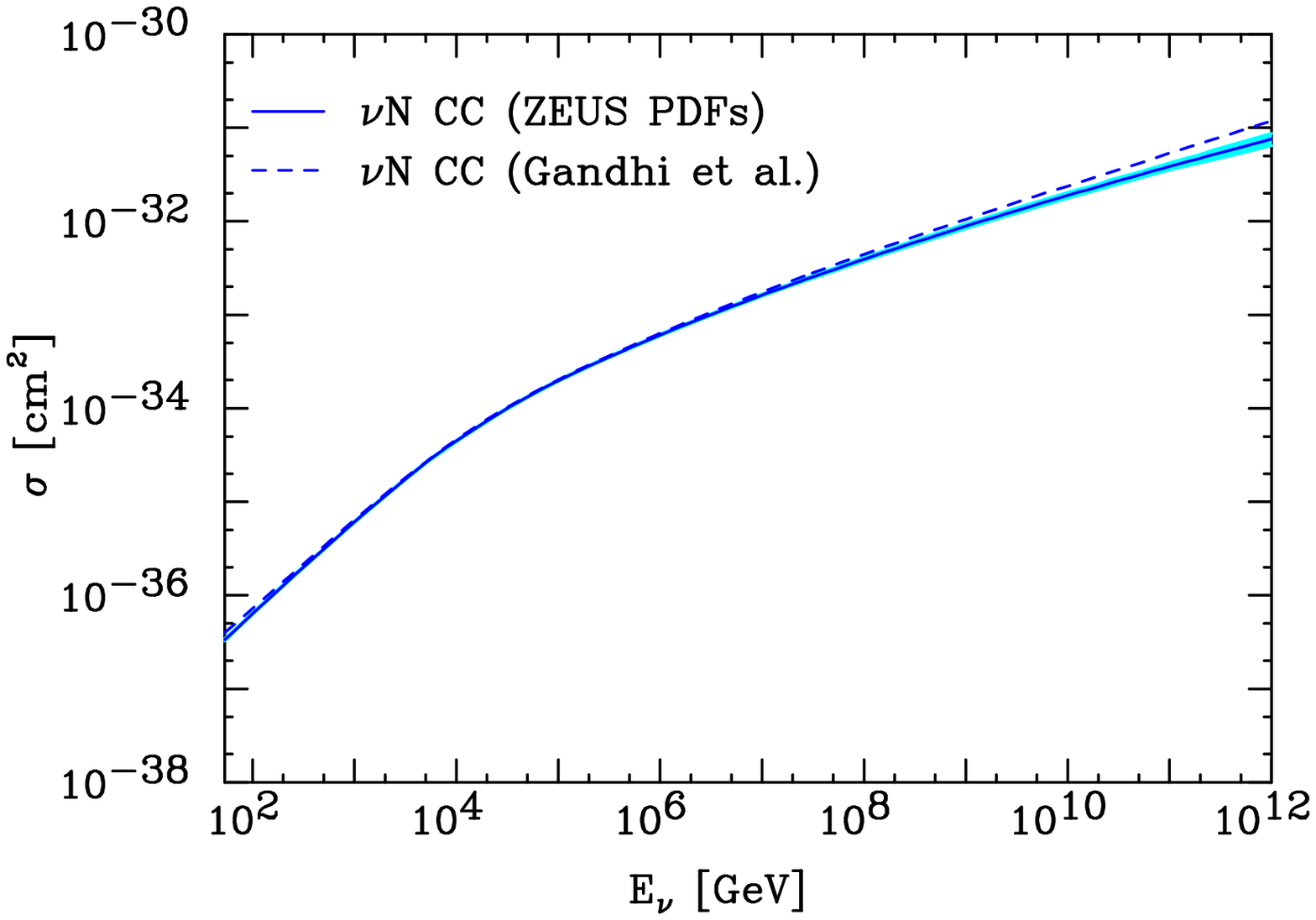,angle=90,width=0.5\textwidth}
\epsfig{figure=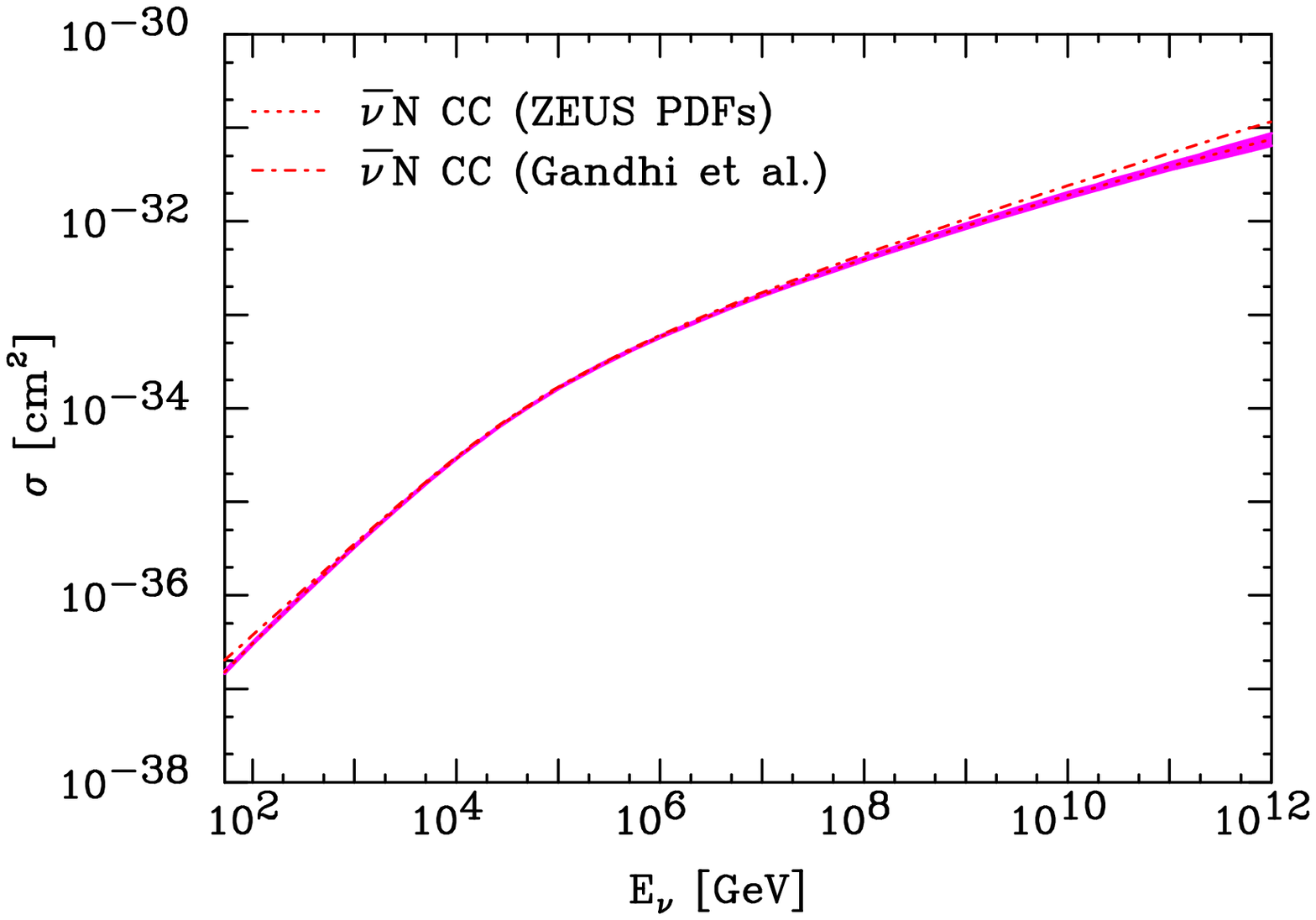,angle=90,width=0.5\textwidth}}
\caption {The total charge-current cross-section at ultra high energies for
  neutrinos and antineutrinos along with the $\pm
  1\sigma$ uncertainties (shaded band) \cite{{CooperSarkar:2007cv}}.}
\label{fig:comparison}
\end{figure}

To estimate detection rates for UHE neutrinos we also need to know the
cross-section for their scattering on nucleons, at energies far beyond
those achievable at terrestrial accelerators. In the framework of the
quark-parton model, such (deep inelastic) scattering accesses very
large values of $Q^2$, the invariant mass of the exchanged vector
boson, and very small values of Bjorken $x$, the fraction of the the
incoming nucleon momentum taken by the struck quark. Thus we need to
extrapolate the experimentally measured parton distribution functions
(PDFs) of the nucleon to the relevant kinematic range using the DGLAP
formalism of perturbative QCD. This is best done using up to date
information from the experiments at HERA, which have accessed the
lowest $x$ and highest $Q^2$ scales to date. We have used the ZEUS-S
global PDF fits \cite{Chekanov:2002pv}, updated to include all the
HERA-I data, at next-to-leading-order and with corrections for heavy
quark thresholds --- the results \cite{CooperSarkar:2007cv} are shown
in Fig.\ref{fig:comparison} along with earlier
values~\cite{Gandhi:1998ri} which were calculated at leading order and
using PDFs which no longer fit modern data. We also provide a measure
of the uncertainties which derive mainly from the correlated
systematic errors of the input data sets. These updated cross-sections
have been used in recent Auger analyses \cite{Abraham:2007rj} and are
being incorporated into ANIS, the MC event generator for neutrino
telescopes \cite{Gazizov:2004va}

There are additional theoretical uncertainties at very high energies
($>10^8$ GeV) since at the very low-$x$ values probed the gluon
density is rising rapidly so it is probably necessary to go beyond the
DGLAP formalism in order to sum $\ln(1/x)$ diagrams, as in the BFKL
formalism. An alternative approach is to consider non-linear terms
which describe gluon recombination as in the `colour glass condensate'
model which has had considerable success in explaining RHIC
data. These non-perturbative effects can {\em reduce} the
cross-section at high energies by a factor of $\sim2-10$. Whether this
is indeed the case can in principle be tested by measuring the zenith
angle dependence of the cosmic UHE neutrino flux. For example in an
air shower array like Auger, the rate of quasi-horizontal events due
to neutrinos interacting in the atmosphere is proportional to the
cross-section, but the rate of Earth-skimming events due to tau
neutrinos interacting in the Earth's crust is approximately
independent of the cross-section (if it is {\em reduced} as above), so
their ratio provides a diagnostic \cite{Anchordoqui:2006ta}. However
the expected low event rates would require much larger detection
volumes than are presently available e.g. a satellite-borne
fluorescence detector like EUSO has been considered
\cite{PalomaresRuiz:2005xw}. Proposed extensions of Cherenkov
detectors like IceCube using radio detection techniques also seem very
promising in this regard.

\newpage
\section{Conclusions}

Ultrahigh energy cosmic neutrinos have not yet been detected but there
is no doubt that they exist in Nature and after years of effort
experiments are approaching the sensitivity at which the ``guaranteed
cosmogenic flux'' should be seen. It has long been recognised that
this would open up a new astronomy and be a decisive step towards
identifying the sources of cosmic rays. It may also be possible using
this free UHE beam of neutrinos to discover new physics both in and
beyond the Standard Model. This is a fertile ground for the meeting of
astrophysics and particle physics and the future indeed looks bright.

\ack{I wish to thank all my colleagues in Auger and IceCube and
  especially my co-authors Luis Anchordoqui, Amanda Cooper-Sarkar, Dan
  Hooper and Andrew Taylor with whom the calculations discussed here
  was carried out. This work was supported by a STFC Senior Fellowship
  (PPA/C506205/1) and the EU network `UniverseNet'
  (MRTN-CT-2006-035863).

\end{document}